\documentstyle[prl,aps,epsfig,multicol]{revtex} 

\input{epsf}

\begin{document}

\draft 

\title{Ground state laser cooling using electromagnetically induced transparency} 

\author{Giovanna Morigi$^1$, J\"urgen Eschner, Christoph H. Keitel$^2$} 
\address{$^1$Max-Planck Institut f\"ur Quantenoptik, D-85748 Garching, Germany\\ Institut f\"ur Experimentalphysik, University of Innsbruck, A-6020 Innsbruck, Austria\\ $^2$Theoretische Quantendynamik, Fakult\"at f\"ur Physik,University of Freiburg, D-79104 Freiburg, Germany} 

\date{\today}

\maketitle 

\begin{abstract}

A laser cooling method for trapped atoms is described which 
achieves ground state cooling by exploiting quantum interference 
in a driven $\Lambda$-shaped arrangement of atomic levels. The 
scheme is technically simpler than existing methods of sideband 
cooling, yet it can be significantly more efficient, in particular 
when several motional modes are involved, and it does not impose 
restrictions on the transition linewidth. We study the full 
quantum mechanical model of the cooling process for one motional 
degree of freedom and show that a rate equation provides a good approximation. 

\end{abstract}



\begin{multicols}{2}

\vspace{0mm} 

Laser cooling has played a central role in the preparation of 
fundamental quantum mechanical atomic systems \cite{Reviews}, for 
example in experiments which study the quantum statistical 
properties of atoms \cite{Qstatistics} or which use trapped ions 
for processing information at the quantum level 
\cite{ExpIons,ExpIonsIbk}. In the latter context, laser cooling of 
trapped ions to the ground state of the confining potential is a 
fundamental step in the preparation of the ion trap quantum 
computer \cite{Qcomputer}. Furthermore, the same techniques that 
allow ground state cooling are at the basis of coherent 
manipulation, i.e.\ gate operations, in quantum computation 
schemes with trapped ions. For both purposes, cooling and gate 
operations, the speed of the manipulation has become an important 
issue \cite{Andrew}, because higher speed means that competing 
heating or decoherence due to coupling to the environment has less 
opportunity to perturb the desired processes. 

Efficient ground state laser cooling of single trapped ions has 
been achieved using two-level sideband cooling 
\cite{Diedrich_und_99:Roos} and Raman sideband cooling 
\cite{Monroe}, and recently these methods have been transferred to 
two ions \cite{Wineland2_und_Rohde} and to atomic gases 
\cite{AtomRamanSB}. These techniques involve laser excitation of 
an atom with two internal levels $|g\rangle$ and $|e\rangle$, 
which in the case of Raman sideband cooling is designed from a 
$\Lambda$-shaped three-level atom by Raman coupling 
\cite{Java,Irene}. Both techniques rely on several conditions 
\cite{Stenholm}: (i) the motional spectrum of the system has 
equidistant levels $|n\rangle$, which is true when the particle 
(or particles) are trapped in a harmonic potential; (ii) the 
amplitude of the oscillations of the trapped particles is much 
smaller than the wavelength of the cooling laser ("Lamb-Dicke 
regime"); (iii) The linewidth $\gamma$ of the internal transition 
is much smaller than the distance between any pair of motional 
energy levels ("strong confinement"). For the case of a single 
particle confined in a harmonic oscillator potential with 
frequency $\nu$, the strong confinement condition is 
$\gamma\ll\nu$. 

Under these conditions it is possible to selectively excite 
sidebands of the optical resonance, i.e.\ transitions 
corresponding to a fixed change of the vibrational quantum number 
$n$ to $n^{\prime}$, by tuning the laser into resonance with that 
transition, while all other transitions are well off resonance and 
thus only negligibly excited. Specifically, for sideband cooling 
transitions $|g,n\rangle \to |e,n-1\rangle$ are induced by tuning 
the laser to $\omega_a - \nu$, i.e.\ to the "red sideband" of the 
bare atomic resonance at frequency $\omega_a$. When a spontaneous 
decay from $|e\rangle$ to $|g\rangle$ takes place or, in the case 
of Raman sideband cooling, when the atom is optically pumped back 
from $|e\rangle$ to $|g\rangle$, this decay occurs with highest 
probability on the transition $|e,n-1\rangle\to |g,n-1\rangle$ due 
to the Lamb-Dicke condition. Thus in one fluorescence cycle the 
system is cooled, on average, by one vibrational quantum. The 
cooling limit is determined by the equilibrium between these 
cooling cycles and heating processes. Heating is induced by 
off-resonant excitation of the $|g,n\rangle \to |e,n\rangle$ 
"carrier" transition followed by a $|e,n\rangle$ to 
$|g,n+1\rangle$ spontaneous emission event, or by excitation of a 
$|g,n\rangle \to |e,n+1\rangle$ "blue sideband" transition. Since 
the selective excitation of the $|g,n\rangle \to |e,n-1\rangle$ 
sideband is at the basis of this technique, this imposes a 
limitation on the intensity of the cooling laser and thus also on 
the cooling speed. In particular, high laser intensity leads to 
increased off-resonant excitation of carrier transitions which 
limits the final ground state occupation of the cooling process. 

In this paper we describe a method for ground state cooling of 
atoms with a multi-level structure which eliminates the carrier 
excitation by electromagnetically induced transparency 
\cite{lots_of_EIT_references}. The technique is based on 
continuous laser excitation and has several advantages over both 
2-level and Raman sideband cooling. Unlike in 2-level sideband 
cooling, no strong confinement is required, instead two 
dipole-allowed transitions are used, neither of which has to 
fulfil the relation $\gamma\ll\nu$. Unlike Raman sideband cooling 
which involves an additional repumping laser, only two lasers are 
needed in our method. Finally, as will be shown, by cancelling the 
carrier transition our scheme provides more efficient ground state 
cooling than sideband cooling methods, in particular for 
simultaneous cooling of several modes of vibration. This work 
extends previous analyses of laser-cooling in a three-level atomic 
system \cite{Java,Irene}, which however focused on different 
cooling mechanisms, as will be discussed below. 

Electromagnetically induced transparency (EIT) arises in three- 
(or multi-) level systems and consists in the cancellation of the 
absorption on one transition induced by simultaneous coherent 
driving of another transition. The phenomenon is also called 
"coherent population trapping" \cite{Knight_Stroud} or "dark 
resonance" and has been demonstrated in many systems 
\cite{EIT_experiments} including single trapped ions 
\cite{PET_dark_res}. It belongs to a large class of quantum 
interference effects in multi-level systems 
\cite{more_quantum_interference} and can be understood as a 
destructive interference of the two pathways to the excited level 
\cite{Cohen}. It is also at the basis of velocity selective 
coherent population trapping (VSCPT), a laser cooling method for 
{\it free} atoms which achieves sub-recoil temperatures 
\cite{VSCPT}. Here, we use this situation to suppress absorption 
on the $|g,n\rangle \to |e,n\rangle$ transition, while enhancing 
the absorption on the $|g,n\rangle \to |e,n-1\rangle$ sideband 
transition, thus decreasing the heating and increasing the cooling 
rate. 

Let us for the moment neglect the motional degrees of freedom and 
consider a 3-level atom with ground state $|g\rangle$, stable or 
metastable state $|r\rangle$ and excited state $|e\rangle$ in 
$\Lambda$-configuration as shown in Fig.\ 1. State $|e\rangle$ has 
linewidth $\gamma$ and is coupled to both $|g\rangle$ and 
$|r\rangle$ by dipole transitions. The transition $|r\rangle\to 
|e\rangle$ is excited by an intense "coupling" laser field of 
frequency $\omega_{r}$, Rabi frequency $\Omega_r$ and detuning 
$\Delta_r= \omega_r -\omega_{re}$, where $\omega_{re}$ is the 
frequency of the bare atomic transition $|r\rangle\to |e\rangle$. 
The absorption spectrum observed by exciting the transition 
$|g\rangle\to |e\rangle$ with another "cooling" laser at frequency 
$\omega_{ge} +\Delta_g$ and Rabi frequency $\Omega_g$ is described 
by a Fano-like profile \cite{Cohen}, whose zero corresponds to the 
case $\Delta_g=\Delta_r$ and which is asymmetric for $\Delta_r\neq 
0$, see Fig.\ 1. The same spectrum describes the rate at which 
photons are scattered from state $|e\rangle$, and one can infer 
from it the cooling effect of the laser excitation on the ion 
\cite{Neuhauser}. 

In the case $\Delta_r > 0$ which is displayed in Fig.\ 1, the two 
components of the absorption spectrum, i.e. the broad resonance at 
$\Delta_g \simeq 0$ with linewidth $\gamma^{\prime\prime} \simeq 
\gamma$ and the narrow resonance at $\Delta_g \simeq \Delta_r$ 
with linewidth $\gamma^{\prime} \ll \gamma$, correspond to the 
dressed states of the system atom + coupling laser \cite{Cohen1}, 
see Fig.\ 1c. These dressed states, and hence the maxima of the 
narrow and broad curve, are shifted from $\Delta_r$ by $+\delta$ 
and $-\Delta_r - \delta$, respectively, with 
\begin{equation} 
\label{ACStark} \delta=(\sqrt{\Delta_r^2+\Omega_r^2}-|\Delta_r|)/2 
\end{equation}
\noindent being the AC Stark shift induced by the coupling laser. 
\begin{center} 
\begin{figure}[tbp] 
\epsfig{file=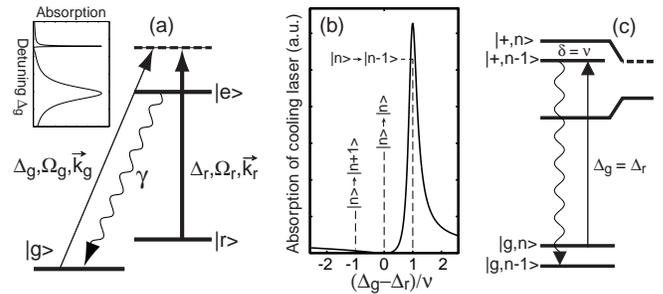,width=0.99\hsize} \vspace{\baselineskip} 
\caption{(a) Levels and transitions of the cooling scheme (found 
in many species used for ion trapping). The inset shows 
schematically the absorption rate on $|g\rangle\to |e\rangle$ when 
the atom is strongly excited above resonance on $|r\rangle \to 
|e\rangle$. (b) Absorption of cooling laser around $\Delta_g = 
\Delta_r$ (solid line) and probabilities of carrier ($|n\rangle 
\to |n\rangle$) and sideband ($|n\rangle \to |n \pm 1\rangle$) 
transitions when $\Delta_g = \Delta_r$ (dashed lines). (c) The 
cooling laser excites resonantly transitions from $|g,n\rangle$ to 
the narrow dressed state \protect\cite{Cohen1} denoted by 
$|+,n-1\rangle$.} 
\end{figure} 
\end{center}

With the harmonic motion taken into account, the zero of the 
Fano-like profile at $\Delta_g=\Delta_r$ corresponds to the 
$|g,n\rangle \to |e,n\rangle$ transition which is therefore 
cancelled. Then, by choosing $\Delta_r>0$ and a suitable Rabi 
frequency $\Omega_r$, the spectrum can be designed such that the 
$|g,n\rangle \to |e,n-1\rangle$ (red) sideband corresponds to the 
maximum of the narrow resonance, whereas the blue sideband falls 
into the region of the spectrum of small excitation probability, 
as shown in Fig.\ 1b. The condition on the laser parameters for 
enhancing the red-sideband absorption while eliminating the 
carrier is therefore: 
\begin{equation} 
\label{settings} 
\Delta_g = \Delta_r \hspace{5mm} ; \hspace{5mm} \delta \simeq \nu 
\end{equation}
\noindent The laser parameters of Eq.\ (\ref{settings}) are easily 
achievable in single ion experiments 
\cite{ExpIonsIbk,Diedrich_und_99:Roos} where typically $\gamma\sim 
2\pi\times$~20~MHz and $\nu\sim 2\pi\times$~1~MHz. 

Note that the detunings are different from Raman sideband cooling 
(RSC) where $\Delta_g=\Delta_r-\nu$. Furthermore, here both lasers 
{\it must} be {\it blue}-detuned from their respective atomic 
resonances, whereas in RSC they can be tuned either both below or 
both above resonance. Moreover, in RSC the bare states 
$|g\rangle$, $|r\rangle$, are coupled under saturation to 
$|e\rangle$ (this is the situation described in \cite{Java,Irene}) 
whereas in the new cooling scheme, both multiple scattering on the 
transition $|r\rangle\to |e\rangle$ and the quantum interference 
at $\Delta_g=\Delta_r$ are crucial for the cooling process, which 
is therefore adequately described by transitions between state 
$|g\rangle$ and the two dressed states, see Fig.\ 1c. 

The mechanism is theoretically modelled as follows. We start with 
the master equation for the full three-level system and one 
motional degree of freedom. In the Lamb-Dicke regime, the master 
equation can be reduced to a rate equation projected on the 
internal state $|g,n\rangle$ provided that $\Omega_g \ll \Omega_r$ 
and that the transition to the narrow dressed state of linewidth 
$\gamma^{\prime}$ is not saturated, i.e.\ $\Omega_g \ll 
\sqrt{\gamma \gamma^{\prime}}$. Then, in second order of the 
expansion in $\Omega_g/ \sqrt{\gamma\gamma^{\prime}}$ the dynamics 
is described by an equation for the populations $P(n)$ of the 
vibrational number states $|n\rangle$ \cite{Lambropoulos}: 

\begin{eqnarray} 
\label{SolLDL} 
\frac{d}{dt}P(n)&=&\eta^2[A_-((n+1)P(n+1)-nP(n))+ \nonumber\\ 
                &+&A_+(nP(n-1)-(n+1)P(n))] 
\end{eqnarray}

\noindent Here, $\eta$ is the Lamb-Dicke parameter, defined as 
$\eta=|\vec{k_g}-\vec{k_r}|a_0$ with $a_0$ rms size of the ground 
state of the harmonic oscillator and $\vec{k_g}$ ($\vec{k_r}$) 
cooling (coupling) laser wave vector \cite{footnoteta}. The 
coefficients $A_{\pm}$ have the form 
\begin{equation} 
\label{Apm} A_{\pm}=\frac{\Omega_g^2}{\gamma} \hspace{1mm} 
\frac{\gamma^2\nu^2} {\gamma^2\nu^2 + 
4\left[\Omega_r^2/4-\nu\left(\nu\mp\Delta\right)\right]^2} 
\end{equation}
\noindent where we have set $\Delta_r=\Delta_g=\Delta$. Note that 
while Eq.\ (\ref{SolLDL}) has the same structure as the rate 
equation that describes cooling of a two-level atom in the 
Lamb-Dicke regime \cite{Stenholm}, the particular form of 
$A_{\pm}$ in Eq.\ (\ref{Apm}) contains the full quantum 
interference around $\Delta_g = \Delta_r$. Solving Eq.\ 
(\ref{SolLDL}) for the steady state value $\langle n_S \rangle$ of 
the mean vibrational quantum number $\langle n \rangle = \sum n 
P(n)$ we get 
\begin{equation} 
\label{Steady} 
\langle n_S \rangle = \frac{A_+}{A_- -A_+} 
= \frac{\gamma^2\nu^2+4[\Omega_r^2/4-\nu(\nu+\Delta)]^2} 
{4\Delta\nu(\Omega_r^2-4\nu^2)} 
\end{equation}
\noindent Eq.\ (\ref{Steady}) has a pole ($A_+=A_-$) at 
$\Delta=0$, where the spectrum is symmetric, and at 
$\Omega_r=2\nu$, where the value of the absorption spectrum at the 
two frequencies $\Delta_g \pm \nu$ is the same. For properly 
chosen parameters, however, a value of $\langle n_S \rangle$ close 
to zero can be reached, as shown in the example of Fig.\ 2. The 
optimum value $\langle n_S \rangle = (\gamma/4\Delta)^2$ is found 
when the second term in the numerator of Eq.\ (\ref{Steady}) 
vanishes, which corresponds precisely to the condition $\delta = 
\nu$ in Eq.\ (\ref{settings}). 
\begin{center} 
\begin{figure}[tbp] 
\epsfig{file=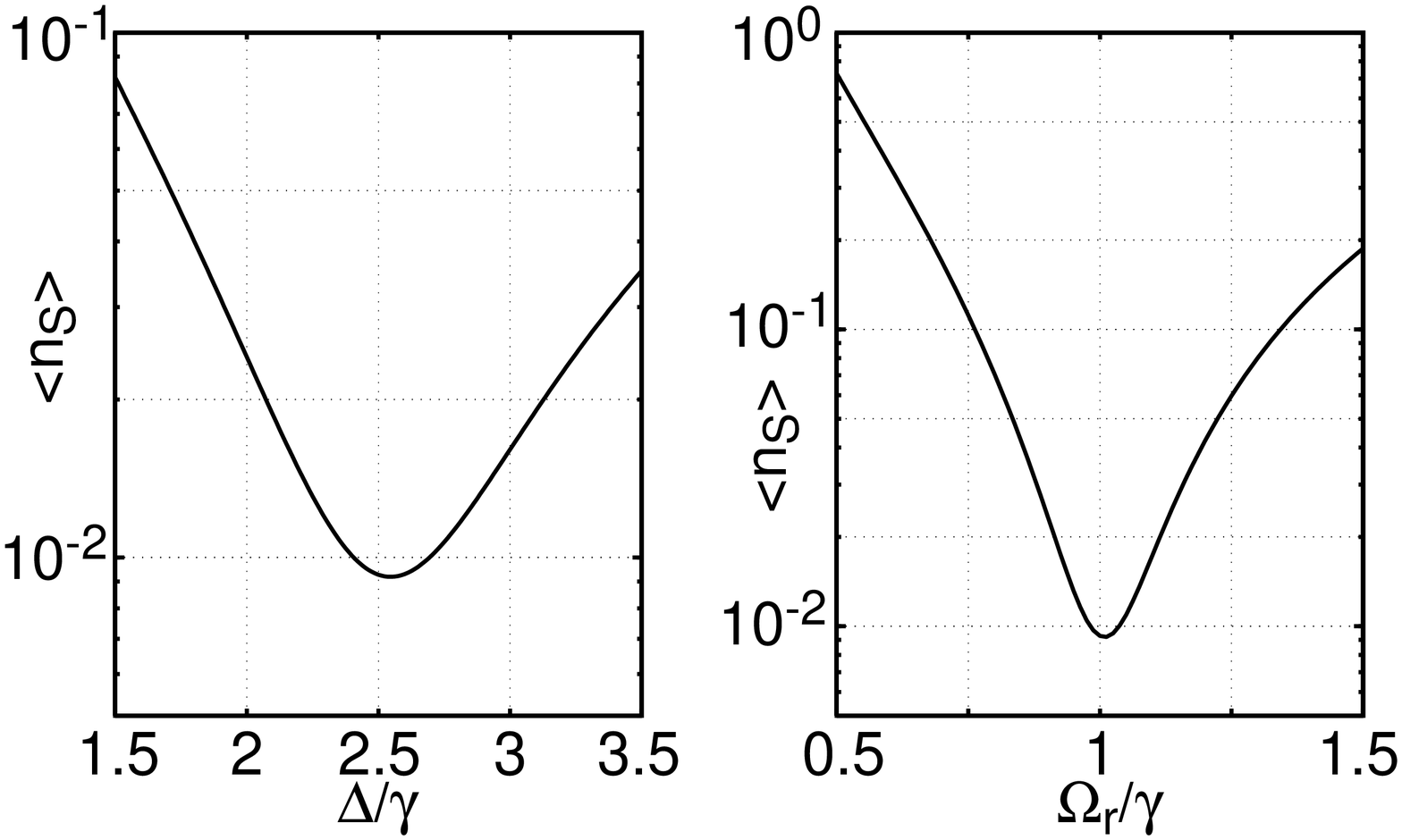,width=0.8\hsize}\vspace{\baselineskip} 
\caption{Cooling limit $\langle n_S\rangle$ (Eq.~(\ref{Steady})) 
as a function of detuning $\Delta/\gamma$ (left) and Rabi 
frequency of the coupling laser $\Omega_r/\gamma$ (right). The 
other parameters are $\nu=\gamma/10$, $\Omega_r=\gamma$ (left), 
and $\Delta=2.5\gamma$ (right). } 
\end{figure} 
\end{center} 

From Eq.\ (\ref{SolLDL}) the time dependence of $\langle n 
\rangle$ follows,
\begin{equation} \label{Evolution} \dot{\langle n \rangle} = 
- \eta^2 (A_- - A_+) \langle n \rangle + \eta^2 A_+ 
\end{equation}
\noindent where $\eta^2 (A_- - A_+)$ is the cooling rate. This 
rate together with $\langle n_S \rangle$ determines the efficiency 
of the cooling technique which is compared to conventional 
sideband cooling below. 

The dynamics of the full system for any set of parameters can be 
calculated with a quantum Monte-Carlo simulation \cite{QMC}. In 
Fig.\ 3 we plot the result of such a calculation and compare it to 
the rate equation solution of Eq.\ (\ref{Evolution}). We see that 
the rate equation provides a good description of the cooling. In 
this example, 99\% occupation of the ground state is achieved. 
\vspace{-5mm}
\begin{center} 
\begin{figure}[tbp] 
\epsfig{file=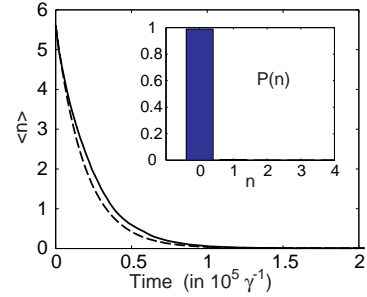,width=0.55\hsize} \vspace{\baselineskip} 
\caption{Onset: $\langle n\rangle$ as a function of time 
calculated with full Monte Carlo simulation (solid line) and rate 
equation (dashed line). Parameters are $\Omega_r=\gamma$, 
$\Omega_g=\gamma/20$, $\nu=\gamma/10$, $\eta=0.145$, 
$\Delta_g=\Delta_r=2.5\gamma$. The atomic parameters are for the 
(S$_{1/2}$, P$_{1/2}$, D$_{3/2}$)-$\Lambda$-system of a Ca$^+$ ion 
\protect\cite{ExpIonsIbk} where $\gamma=2\pi\times20$~MHz. Inset: 
Steady state distribution $P(n)$.} 
\end{figure} 
\end{center}

In order to compare our EIT-cooling scheme with conventional 
2-level sideband cooling, we use the results of the respective 
rate equations for both schemes and we allow for an angle $\phi$ 
between the motional axis and the direction of the laser beam (of 
$\vec{k_g}-\vec{k_r}$ in the case of EIT-cooling). We denote by 
$A_{\pm}^{\rm SC}$ the coefficients for sideband cooling 
corresponding to Eq.\ (\ref{Apm}), c.f.\ \cite{Stenholm}. Assuming 
a linewidth $\gamma_{\rm SC} = \gamma^{\prime}$  for the sideband 
cooling transition, and the same degree of saturation in both 
schemes, i.e.\ $(\Omega_{\rm SC}/\gamma_{\rm SC})^2 = 
\Omega_g^2/\gamma\gamma^{\prime}$, the following relation holds: 
\begin{equation} 
\label{rates} A_{\pm}^{\rm SC} = A_{\pm} + \frac {\Omega_{\rm 
SC}^2} {\gamma_{\rm SC}} \left(\frac {\alpha} {\cos^2\phi}\right) 
\frac{\gamma^{2}_{\rm SC}}{\gamma^{2}_{\rm SC}+4\nu^2} 
\end{equation} 
\noindent Here we have used that $\gamma^{\prime} \approx \gamma 
\frac{\nu}{\Delta}$ for our conditions $\delta = \nu \ll 
\Delta,\Omega_r$ \cite{Cohen1}, and $\alpha=\int_{-1}^1du N(u) 
u^2$ with $N(\cos\vartheta)$ being the azimuthal dipole pattern of 
spontaneous emission on the sideband cooling transition. The 
additive term in Eq.\ (\ref{rates}) highlights that the difference 
between sideband cooling and EIT-cooling lies in the heating that 
accompanies carrier absorption, which in EIT-cooling is cancelled. 
This leads to the remarkable result that in contrast to any other 
cooling scheme, in EIT-cooling the theoretical limit $\langle n_S 
\rangle$ does not depend on the angle between the direction of the 
laser beams (of $\vec{k_g}-\vec{k_r}$) and the motional axis. For 
the typical condition of sideband cooling $\gamma_{\rm SC} \ll 
\nu$ we find, using Eq.\ (\ref{Steady}), 
 
\begin{equation} 
\langle n_S \rangle_{\rm EIT} = \frac {\langle n_S \rangle_{\rm 
SC}} { 1 + \frac {4\alpha} {\cos^2\phi}} 
\end{equation} 

\noindent Thus, given the same cooling rate, $\langle n_S 
\rangle_{\rm EIT}<\langle n_S \rangle_{\rm SC}$. The factor 
between the two steady state values is $\frac{5}{29}$ for all 
three motional degrees of freedom, assuming the ion is cooled in 
three dimensions by a single pair of laser beams, 
$\alpha=\frac{2}{5}$ \cite{Stenholm}, and $\vec{k_g}-\vec{k_r}$ 
has the same angle with all axes of the trap. It becomes even 
smaller when $\vec{k_g}-\vec{k_r}$ is at a large angle with the 
motional axis. This makes EIT-cooling a significantly more 
efficient technique in many typical experimental situations 
\cite{Ferdi}. Moreover, numerical studies show that the efficiency 
can be increased further if the cooling laser is tuned not exactly 
to the dark resonance but slightly above such that the combined 
heating of carrier and blue sideband transitions is minimised. 

In conclusion, we have presented a laser cooling technique for 
trapped particles which exploits quantum interference, or 
electromagnetically induced transparency, in a 3-level atom. By 
appropriately designing the absorption profile with a strong 
coupling laser, the cooling transitions induced by a cooling laser 
are enhanced while heating by resonant absorption is suppressed. 
The method is not based on the strong confinement condition and 
requires only two continuous lasers. We have derived a simple 
model for describing the cooling process and shown that it is in 
good agreement with a full quantum Monte Carlo treatment. With the 
same cooling rate as in conventional sideband cooling, much higher 
ground state occupation is achieved by our method, in particular 
if three-dimensional cooling is considered. The technical 
requirements, two lasers with a well-controlled frequency 
difference, are met by most existing single ion experiments, and 
they are less stringent than for both Raman sideband cooling and 
ordinary 2-level sideband cooling. Furthermore, the method is 
insensitive to laser frequency fluctuations as long as the laser 
linewidth is small compared to the trap frequency. 

Simultaneous cooling in three dimensions can be achieved with this 
method if the trap frequencies along the axes are similar, so that 
the red sideband for each oscillator falls into the neighbourhood 
of the maximum of the narrow resonance. Similarly, the method can 
be applied to simultaneously cool several axial modes of an N-ion 
crystal in a linear trap, or atoms in anharmonic traps, where the 
energies of the motional states are not equidistant. An extension 
would be to use an atom with more than three levels, where 
multiple dark resonances occur (see, e.g., \cite{PET_dark_res}) 
and to design the absorption spectrum such that both the carrier 
and the blue sideband transition vanish. 

{\it Note}: Following the ideas of this proposal, the method has 
meanwhile been experimentally demonstrated \cite{EITexp}. 

We thank J.\ I.\ Cirac, P.\ Lambropoulos, D.\ Leibfried, C.\ Roos, 
F.\ Schmidt-Kaler, H.\ Walther, and P.\ Zoller for many 
stimulating discussions, and acknowledge support by the European 
Commission (TMR networks ERB-FMRX-CT96-0077 and 
ERB-FMRX-CT96-0087; Marie Curie Program) and the German Science 
Foundation (SFB 276). CHK is thankful for hospitality at Innsbruck 
University.

\end{multicols}

\end{document}